\documentclass[aps,pra,twocolumn,showpacs,preprintnumbers,amsmath,amssymb]{revtex4}

\usepackage{graphicx, graphics}
\usepackage{psfrag}
\usepackage{bm}
\usepackage[below]{placeins}
\usepackage{array}
\usepackage[usenames]{color}
\usepackage{textcomp}
\begin{document}

\title{Anisotropic pair-superfluidity of trapped two-component Bose gases}
\author{Yongqiang Li$^{1,2}$, Liang He$^{1}$ and Walter Hofstetter$^{1}$}
\address{$^{1}$Institut f\"ur Theoretische Physik, Goethe-Universit\"at, 60438 Frankfurt am Main, Germany \\
$^{2}$Department of Physics, National University of Defense Technology, Changsha 410073, P. R. China}
\date{\today}

\begin{abstract}
We theoretically investigate the pair-superfluid phase of two-component ultracold gases with negative inter-species interactions in an optical lattice. We establish the phase diagram for filling $n=1$ at zero and finite temperature, by applying Bosonic Dynamical Mean-Field Theory, and confirm the stability of pair-superfluidity for asymmetric hopping of the two species. While the pair superfluid is found to be robust in the presence
of a harmonic trap, we observe that it is destroyed already by a small population imbalance of the two species.
\end{abstract}

\pacs{03.75.Kk, 03.75.Mn, 67.85.-d, 67.85.Hj}

\maketitle

\section{Introduction}
Ultracold gases in an optical lattice are promising as flexible quantum simulators for the study of quantum phases which are not easily accessible in condensed-matter physics~\cite{I. Bloch_2008}. Experimental realization of the superfluid-to-Mott transition has paved the way to studies of strongly interacting Bose or Fermi gases in optical lattices~\cite{I. Bloch_2002}. Recently, Bose-Bose mixtures of $^{87}$Rb and $^{41}$K have been produced and loaded into an optical lattice \cite{J. Catani_2008}, which provide a new playground for investigating the interplay between kinetic energy, interaction and the spin degree of freedom. In a further line of investigation, a Bose-Bose mixture of two
hyperfine states of $^{87}$Rb has been used as a spin-gradient thermometer for measuring the temperature of ultracold gases in an optical lattice~\cite {W. Ketterle_2009,W. Ketterle_2010}. Moreover, the effect of a second species on bosonic superfluidity in an optical lattice has been studied~\cite {B. Gadway_2010}. In these mixtures of different species or different hyperfine states, the most fundamental physics associated with quantum magnetism and the spin degree of freedom can be explored~\cite{M. Lewenstein_science}.

Recently, it was found that a three-body hard-core constraint can stabilize a system of single-component bosons with attractive two-body interactions in an optical lattice~\cite{P. Zoller_2009}, and numerical simulations have been
performed to study properties of this system~\cite{M. Yang_2010, M. Yang_2011, S. Wessel_2011}. For two-component bosons with attractive inter-species interactions in an optical lattice, one interesting ground state is the pair-superfluid phase (PSF), which has been predicted and studied theoretically in several papers~\cite{A. Kuklov_1, A. Arguelles_2007, S.
Guertler_2008, L. Mathey_2009, A. Hu_2009, C. Menotti_2010, M.
Iskin_2010, P. Chen_2010}. Qualitatively, the PSF can be viewed as a condensate of bosonic pairs of different species or hyperfine states due to a second-order hopping of the pairs, but with a strongly suppressed first-order tunneling of single
atoms. For hard-core bosons with total filling $n=1$, the PSF for
attractive interactions is equivalent to the XY-ferromagnetic
phase (or super-counter-fluid state, denoted as SCF) for repulsive
interactions, i.e. the PSF consists of particle-particle pairs of
different species while the XY-ferromagnetic phase is composed of particle-hole pairs. It is apparent that the
PSF can only exist at very low temperature compared to the critical temperature of quantum magnetic phases. The latter are also governed by
second-order tunneling processes, leading to an effective spin-exchange coupling~\cite{A. Kuklov_2003, E. Altman_2003}, which has already been observed for a double-well system~\cite{S. Foelling_2007, S. Trotzky_2008}. At the current stage, it is experimentally challenging to
reach the critical temperatures of quantum magnetic phases and the PSF~\cite{W. Ketterle_2010}. It is expected that these phases can be detected via momentum-space correlations observed in time-of-flight measurements~\cite{C. Menotti_2010, U. Shrestha_2010} or by optical microscopy with single-site resolution~\cite{W. Bakr_2009, J. Sherson_2010}.

Previous studies of the PSF in a two- or three-dimensional optical lattice mostly focus on symmetric parameters for the two species (for an exception in~\cite{A. Kuklov_1, M. Iskin_2010}), since this is the most
favorable condition for the PSF~\cite{C. Menotti_2010}. However,
there is still a lack of detailed quantitative investigations of the PSF for the homogeneous system with asymmetric hopping amplitudes of the two species, for the trapped system, and for imbalanced mixtures of the two species. Here we investigate the properties of the PSF of two-component ultracold bosons with attractive interspecies interaction, both in a homogeneous
and a trapped optical lattice. This system can for sufficiently low filling be well described by a
single-band Bose-Hubbard model with pure onsite interaction. We investigate the homogeneous system by means of bosonic dynamical
mean field theory (BDMFT), which is a non-perturbative approach towards strongly-correlated bosonic systems~\cite{BDMFT}, and
the trapped system by real-space bosonic dynamical mean field theory (RBDMFT)~\cite{RBDMFT}, which includes arbitrary inhomogeneity such as a harmonic trap. For the homogeneous system, we focus on the phase diagram with filling
number $n=1$. In particular, we also present the phase diagram
for the experimentally realized heteronuclear $^{87}$Rb - $^{41}$K
mixtures, where double-species Bose-Einstein
condensates with negative inter-species interactions have been observed~\cite{M. Inguscio_2008}. For the trapped Bose-Bose
mixture we study the coexistence of Mott insulator, superfluid and PSF.

The paper is organized as follows: in section II we give a detailed description of the Bose-Hubbard model and the BDMFT approach.
Section III covers our results on the homogeneous Bose-Bose mixture in an optical lattice and the effect of the trap. We conclude in Section V.

\FloatBarrier
\section{Model and method}
We consider a two-component bosonic mixture loaded into a 2D or 3D
optical lattice. In experiments this mixture could consist of two different species, e.g $^{87}$Rb and $^{41}$K
as in \cite{J. Catani_2008} or two different hyperfine states of a single isotope, e.g $^{87}$Rb as in \cite{W. Ketterle_2009}. Besides the
optical lattice, we also impose an external harmonic trapping
potential which introduces inhomogeneity in the system. For sufficiently low filling, the whole
system can be effectively described by a two-component inhomogeneous
Bose-Hubbard model within the single-band approximation
\begin{eqnarray}\label{Hamil}
 \mathcal{H}=&-& \sum_{\stackrel{<i,j>}{\nu=b,d}} t_\nu (b^\dagger_{i\nu}b_{j\nu}+h.c)+\frac{1}{2}\sum_{i,\lambda\nu} U_{\lambda\nu} \hat{n}_{i,\lambda}
(\hat{n}_{i\lambda}-\delta_{\lambda\nu}\nonumber) \\
&+&\sum_{i,\nu=b,d} (V_i-\mu_\nu)\hat{n}_{i\nu}
\end{eqnarray}
where $\langle i,j\rangle$ denotes summation over nearest
neighbor sites $i,j$, and the two bosonic species are labeled
by the index $\lambda (\nu)=b, d$. $\hat{b}^{\dagger}_{i\nu}$ ($\hat{b}_{i\nu}$)
denotes the bosonic creation (annihilation) operator for species $\nu$
at site $i$ and $\hat{n}_{i,\nu}=\hat{b}^\dagger_{i\nu}\hat{b}_{i\nu}$ the corresponding local
density. Due to possibly different masses or a spin-dependent optical lattice, these two species generally hop with non-equal amplitudes
$t_b$ and $t_d$. $U_{\lambda\nu}$ denotes the inter- and intra-species interactions, which can be tuned via Feshbach resonances or spin-dependent lattices \cite{ A. Widera_2004, B. Gadway_2010}. $\mu_\nu$ is the global chemical potential for the two bosonic species, and $V_i \equiv V_0 \, r^2_i $ denotes the external harmonic trap, where $V_0$ is the strength of the harmonic trap and $r_i$ is the distance from the trap center.

To investigate properties of the system, we apply BDMFT for a homogeneous lattice~\cite{BDMFT}, and its real-space generalization (RBDMFT) for the
trapped system~\cite{RBDMFT}. Within RBDMFT, the Hamiltonian~(\ref{Hamil}) is mapped onto a set of individual single-site
problems, and the physical properties on each lattice site are determined by a local effective action which is captured by an effective Anderson impurity model~\cite{BDMFT} solved by exact diagonalization (ED)~\cite{BDMFT,M. Caffarel_1994}. RBDMFT fully captures the inhomogeneity of the system and is capable of providing an accurate and non-perturbative description of quantum phases and their excitations. The detailed formalism is presented in~\cite{RBDMFT}.

To include the effect of spatial inhomogeneity, we also employ an LDA approximation combined with single-site BDMFT. The
advantage of this approach is the larger system size accessible. Within LDA+BDMFT, the local chemical potential for each species is
set to $\mu_\nu(r)=\mu_\nu-V(r)$. In this work, we apply both RBDMFT and
LDA+BDMFT to the 2D trapped square lattice, and only LDA+BDMFT to the 3D
trapped cubic lattice.

\FloatBarrier
\section {Results}
In this section, we will investigate properties of two-component
bosonic mixtures with negative inter-species interactions in a
homogeneous 3D optical lattice at zero and finite temperature, and
also in harmonically trapped, inhomogeneous systems both in 2D and 3D. For the homogeneous system, we will study the
stability of the PSF against asymmetric hopping and finite temperature.
In the presence of an external harmonic trap, we will investigate finite size effects of the PSF both in 2D and 3D within RBDMFT
and LDA+BDMFT. We choose the absolute value of the inter-species
interaction $U\equiv U_{bd}$ as our unit of energy in the following.
\begin{figure}[h]
\vspace{-13mm}
\includegraphics*[width=3.8in]{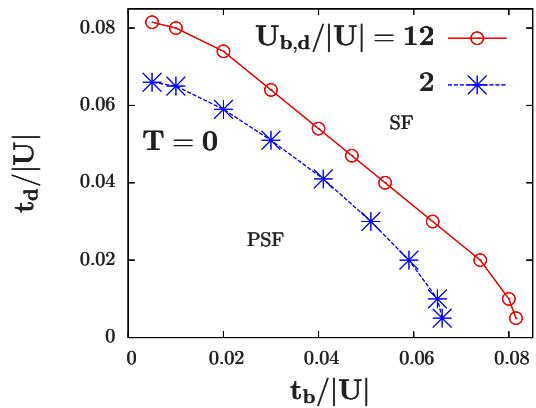}
\vspace{-5mm}
\caption{Zero-temperature phase diagram for two-component bosons with attractive inter-species interaction in
a 3D cubic lattice. The total filling is $n=1$ with $n_b=n_d=0.5$.}\label{phase_diagram}
\end{figure}
\subsection{Bose-Bose mixture on a 3D cubic lattice}
\begin{figure}[h]
\vspace{-12mm}
\begin{tabular}{ c }
\includegraphics*[width=3.8in]{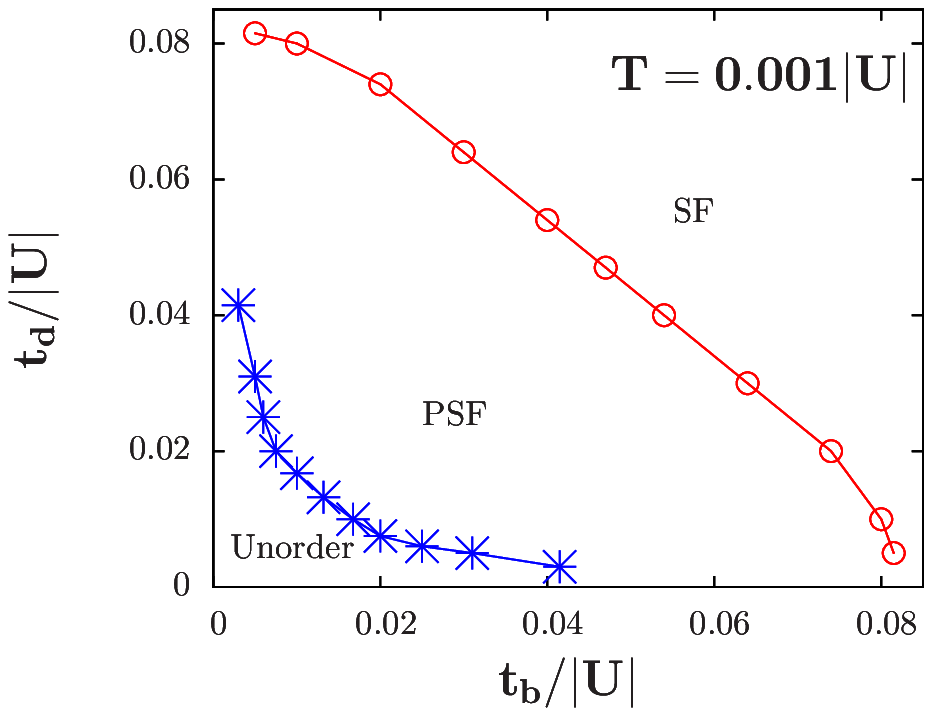}
\vspace{-10mm}
\\
\includegraphics*[width=3.8in]{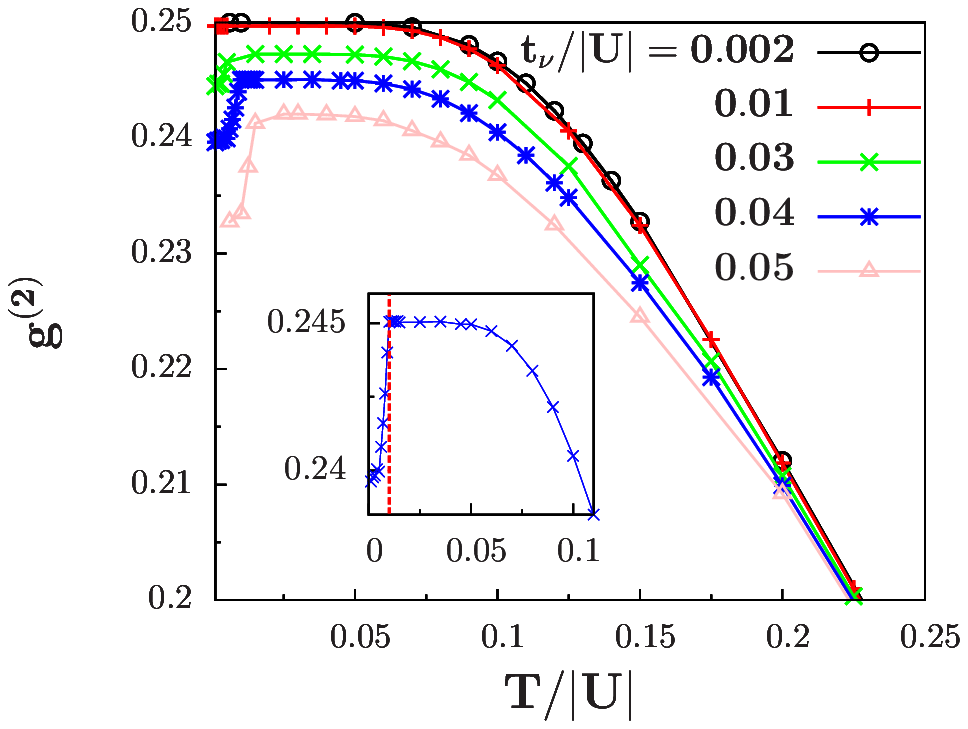}
\end{tabular}
\vspace{-5mm}
\caption{{\bf Upper}: Finite-temperature phase diagram for two-component bosons
in a 3D cubic lattice ($T=0.001|U|$). The interactions are set to $U_b=U_d=12|U|$ and the
total filling is $n=1$ with $n_b=n_d=0.5$. {\bf Lower}: Local density-density correlator $g^{(2)}$ as a function of temperature for different hopping values $t_b=t_d$. Inset: Zoom of main figure for $t_\nu=0.04|U|$ ($\nu=b,d$), where the red dashed line shows the disappearance of the PSF correlator $\phi_{bd}$.}\label{phase_diagram_T}
\end{figure}
We first investigate Bose-Bose mixtures with negative inter-species interaction $U<0$ in a 3D cubic optical lattice.
The system is unstable for $|U|>U_{b,d}$, since the strongly attractive inter-species interaction cannot be compensated by repulsive
intra-species interactions, leading to a collapse of the system. Here we will demonstrate the
stability of the PSF in the regime $|U|<U_{b,d}$. We explore the zero-temperature phase diagram with
asymmetric hopping parameters at total filling $n=1$ (with $n_b=n_d=0.5$) for different interactions, as shown in Fig.~\ref{phase_diagram}. We observe two different phases: the PSF with $\langle b \rangle =\langle d \rangle =0$ but $\phi_{bd}\equiv \langle bd \rangle
- \langle b \rangle \langle d \rangle \neq 0 $, and the superfluid phase (SF) with $\langle b \rangle, \langle d \rangle>0$. In particular, we confirm the existence of a PSF for asymmetric hopping amplitudes $t_b \neq t_d$. In the regime of weak hopping for both species, first-order tunneling is suppressed by the strong interactions. Formation of bosonic pairs between
different species can thus be energetically favored and will typically
compete with single-species condensation, since the
bosonic pairs can hop via second-order tunneling.
As a result, the PSF will have a non-vanishing order parameter $\phi_{bd}$ but vanishing superfluid order $\langle b \rangle$ and $\langle d \rangle$. On the other hand, when both species acquire large hopping, a superfluid phase will appear
with $\langle b \rangle
> 0$ and $\langle d \rangle > 0$. Note that, when the
intra-species interaction $U_{b,d}$ decreases from $U_b=U_d=12|U|$
to $2|U|$, the PSF shrinks due to the decrease of the effective
pair-tunneling amplitudes. We remark here that the transition from a PSF to a superfluid
phase for symmetric hopping amplitudes occurs at the same value of $t_\nu/U_{bd}$
as that from the XY-ferromagnetic to the superfluid phase in the corresponding system with $U_{bd}>0$ of equal magnitude. In contrast to the XY-ferromagnetic phase, however, the PSF also exists for non-integer filling.

The effect of finite temperature is shown in Fig. \ref{phase_diagram_T}.
Generally, the PSF is sensitive to temperature, since the pairs are formed
in the weak hopping regime and their coherence can be easily destroyed by thermal
fluctuations, due to their small effective tunneling of order $O$($t^2/U$). At finite
temperature, the PSF regime shrinks in the weak hopping regime in favor of developing a new unordered
phase with vanishing values for $\phi_{bd}$, $\langle b \rangle$ and $\langle d \rangle$.
To further understand this unordered phase,
we calculate the dependence of the local density-density correlator $g^{(2)}\equiv \langle n_b n_d\rangle -\langle n_b\rangle \langle n_d\rangle$ on temperature, as shown in the lower panel of Fig.~\ref{phase_diagram_T}. We observe that $g^{(2)}$ starts to decrease noticeably only above temperatures of the order of $10^{-1}|U|$, which indicates that local pairs still exist below this temperature. We therefore conclude that the unordered phase, shown in the upper panel of Fig.~\ref{phase_diagram_T}, consists of non-coherent pairs of different species.
In sufficiently deep optical lattices, these pairs are localized, while for larger hopping the pairs delocalize over the whole lattice. As a result, the local density-density correlator decreases as a function of $t_{b,d}$, as shown in the lower panel of Fig.~\ref{phase_diagram_T}. Another interesting feature of the temperature dependence of $g^{(2)}$ is the increasing (non-monotonic) trend at low temperatures, since thermal fluctuations first localize and then break the pairs. We remark here that the temperature regime of non-condensed pairs ($\approx 0.1|U|$) is experimentally accessible~\cite{cooling}, and could be detected via radio frequency spectroscopy~\cite{RF}.
Note that the unordered phase observed here is qualitatively different from that melted from the
XY-ferromagnetic phase, since it can also exist for non-integer filling. We also observe that the (single-particle) superfluid
phase remains almost unchanged for the temperature considered here.

\begin{figure}[h]
\vspace {-12mm}
\includegraphics*[width=3.7in]{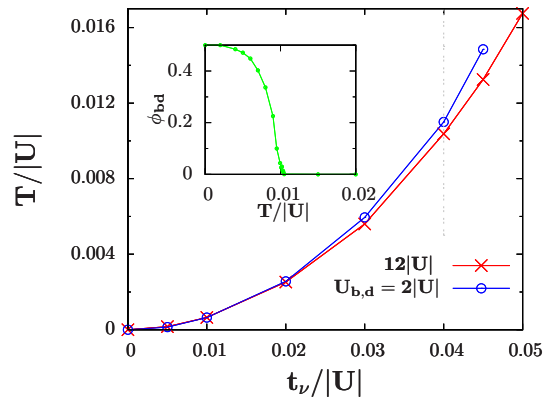}
\vspace{-10pt} \caption{Critical temperature of the PSF as a function
of hopping amplitudes $t_b=t_d$ on a 3D cubic lattice
with total filling $n=1$. Inset: melting of the PSF vs.~temperature
along the vertical dashed line with hopping amplitudes
$t_b=t_d=0.04\,|U|$.}\label{melting}
\end{figure}
One crucial question regarding the observation of the PSF is how fragile
it is against finite-temperature effects. To address this issue,
Fig.~\ref{melting} shows $T_c$ as a function of the hopping
amplitudes $t_b=t_d$ at different interactions. We notice that
$T_c$ rises as the hopping amplitudes increase, due to the growing
second-order tunneling which stabilizes long-range order. The
inset of Fig.~\ref{melting} shows the temperature dependence of
$\phi_{bd}$. It indicates a second-order phase transition
from the PSF to the unordered phase. We also observe that the
critical temperatures for the PSF shown here are comparable to those of the XY-ferromagnetic phase \cite{RBDMFT} and
notably smaller than the coldest temperatures which have been
measured in most experiments until now, with the exception of the MIT group where temperatures as low as
350pK ($\approx 0.01U_{bd}$ with $t_b/U_{bd}\approx 0.029$) have been achieved~\cite{W. Ketterle_2010}.

\begin{figure}[h]
\includegraphics*[width=3in]{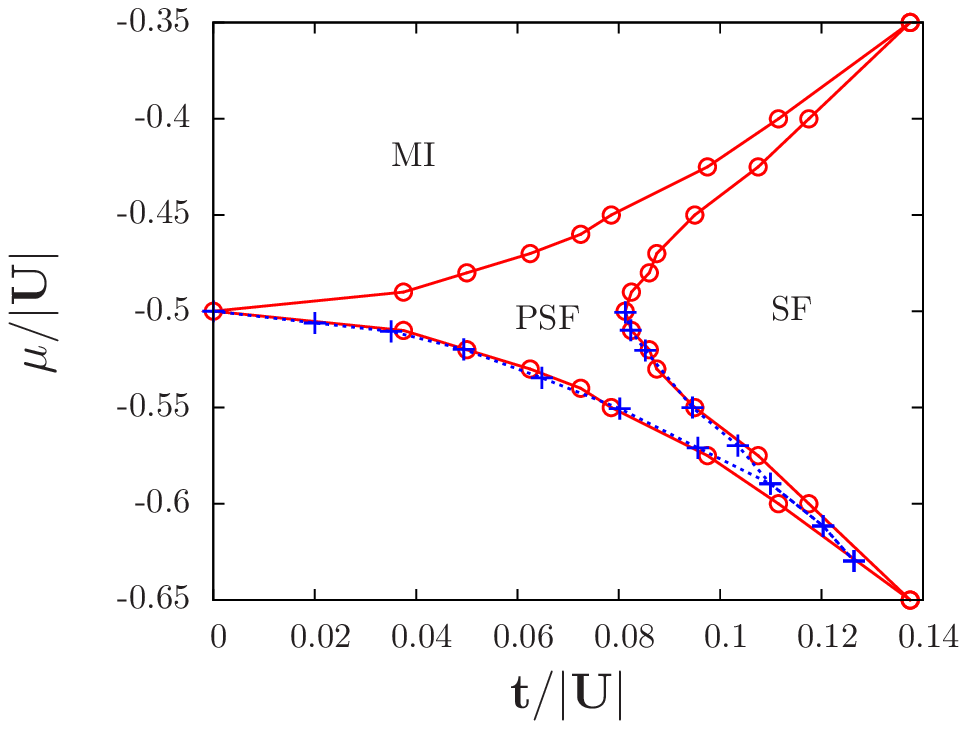}
\includegraphics*[width=3in]{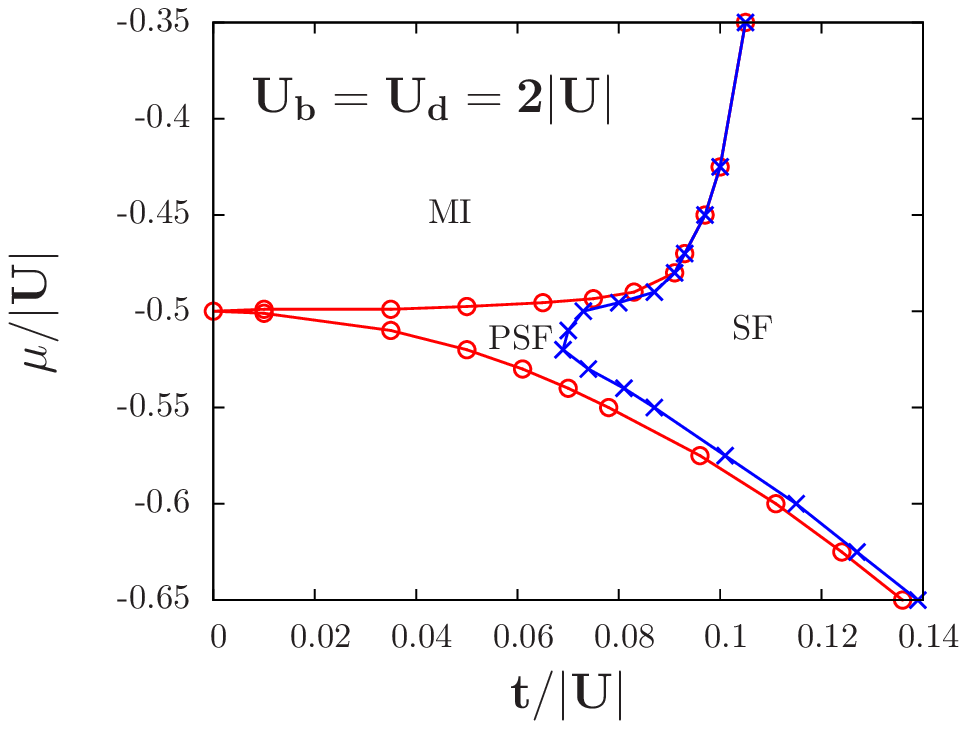}
\caption{{\bf Upper}: Comparison of the zero-temperature phase diagram for
two-component hard-core bosons with the one obtained by a tensor-product-state
approximation \cite{P. Chen_2010} for a square lattice with
symmetric parameters: $t=t_b=t_d$, $\mu=\mu_b=\mu_d$. The red solid lines
are phase boundaries obtained by BDMFT for $U_b=U_d=200|U|$, while the blue
dashed lines are the results of the tensor-product-state approximation. {\bf Lower}:  Zero-temperature phase diagram for
two-component soft-core bosons with $U_b=U_d=2|U|$ obtained via BDMFT.}\label{phase_diagram-hard-core}
\end{figure}
To verify the validity of the BDMFT results, comparison has been made with a hard-core boson model solved by a tensor-product-state approximation \cite{P. Chen_2010}, as shown in Fig.~\ref{phase_diagram-hard-core}. We find excellent agreement between the two methods.
We also plot the phase diagram for soft-core bosons ($U_b=U_d=2|U|$), and observe that in this case the phase boundary between the PSF and the superfluid phase is shifted to lower hopping values.

\FloatBarrier
\subsection{Rubidium-potassium mixture}
\begin{figure}[h]
\includegraphics*[width=\linewidth]{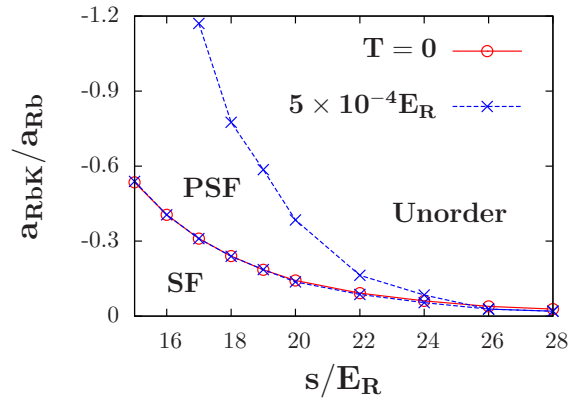}
\caption{Phase diagram for a mixture of $^{87}$Rb and $^{41}$K in a 3D cubic lattice as a function of lattice depth $s$ and Rb-K scattering length.
The total filling is $n=1$ with $n_b=n_d=0.5$.}\label{RK_mixture}
\end{figure}
Our investigations have so far focused on symmetric interactions $U_b=U_d$, which is a good approximation for mixtures of hyperfine states of $^{87}$Rb~\cite{W. Ketterle_2009}. However, this symmetry is not present for mixtures of $^{87}$Rb and $^{40}$K, where a negative inter-species interaction has been achieved via a Feshbach resonance
\cite{M. Inguscio_2008}. Here we consider a mixture of
$^{87}$Rb and $^{41}$K loaded into a 3D cubic lattice
with wavelength $\lambda=757$ nm, which yields equal dimensionless
lattice strength $s$ for the two species. Due to the different masses, the ratio of the intra-species interaction strengths is
then fixed to $U_{\rm Rb}/U_{\rm K}=m_{\rm K}a_{\rm Rb}/m_{\rm Rb}a_{\rm K}\approx0.72$ and the
ratio of the hopping amplitudes to $t_{\rm Rb}/t_{\rm K}\approx0.47$, where $E_{\rm R}$ is the recoil energy.

Now we explore the phase diagram of $^{87}$Rb and $^{41}$K
mixtures in a 3D cubic lattice and make predictions for ongoing experiments. Since the depth $s$ of the optical lattice
and the inter-species scattering length $a_{\rm RbK}$ are tunable
with high accuracy, we show in Fig.~\ref{RK_mixture} the phase diagram in
the $a_{\rm RbK}$-$s$ plane for total filling $n=1$ ($n_b=n_d$) at zero and finite temperatures.
At zero temperature, two phases appear: superfluid
and PSF. When the scattering length $a_{\rm RbK}$ is small, the system is in a superfluid phase for a shallow lattice. When the depth
of the lattice is increased, the ratio of $t_{\rm Rb}/U_{\rm Rb}$ decreases, resulting in a strong suppression of first-order tunneling. The dominant process will then be hopping of composite pairs. At finite temperature, the PSF can be easily
destroyed by thermal fluctuations which induce a second-order phase transition into a unordered phase with $\phi_{bd}=0$. Since the PSF is formed via second-order tunneling and the corresponding energy scale is very small, for the parameters chosen here, this transition already occurs at a low temperature $T=0.0005 E_R$. On the other hand we observe that at the same temperature the superfluid phase is very stable and the phase boundary between superfluid and PSF is almost unchanged.

\begin{figure}[h]
\vspace{-30pt}
\begin{center}
\includegraphics*[width=\linewidth]{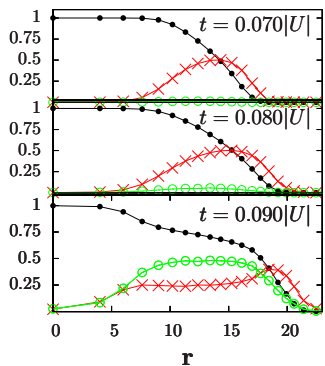}
\end{center}
\vspace{-20pt}
\caption{Density distributions $n_b$ (black line), order
parameters $\phi_b$ (green line) and PSF correlator $\phi_{bd} $ (red line) vs. radial distance $r$ for different
hopping amplitudes at zero temperature in a 2D square lattice, obtained by
RBDMFT. The interactions are $U_b=U_d=2|U|$, hopping amplitudes
$t=t_b=t_d$ and harmonic trap $V_0=0.0002|U|$.}\label{2D_density_trap}
\end{figure}
\FloatBarrier
\subsection{Trapped Bose-Bose mixtures in 2D and 3D lattices}
In this section, we simulate the two-component bosonic system in
both 2D and 3D in the presence of a harmonic trap, as is relevant for most experiments. In particular, we investigate
the stability of the PSF in the trapped system. Here we choose a
$41\times41$ square lattice for the 2D case and a $41\times41\times41$ cubic
lattice for the 3D case. In 2D, in our simulations we apply both RBDMFT and
BDMFT+LDA, while in 3D we only use BDMFT+LDA due to its lower computational effort.

\FloatBarrier
\subsubsection{Balanced mixture}
Fig.~\ref{2D_density_trap} shows the density distributions
$n_b$, order parameter $\phi_b$ and correlator $\phi_{bd}$ for the PSF vs. radius $r$ at different hopping amplitudes
in a trapped 2D optical lattice. Since the PSF is stabilized only
within a narrow region for the symmetric parameters (see Fig.~\ref{phase_diagram-hard-core}), the harmonic trap should be very shallow and the
hopping amplitudes need to be fine-tuned.
Otherwise, the system will go through a phase transition directly
from a Mott-insulating to a superfluid phase. Here we choose completely symmetric parameters: $t=t_b=t_d$ and $U_b=U_d=2|U|$ with balanced filling for the two components. Therefore only one value for $n_{b,d}$ and $\phi_{b,d}$ respectively is shown in Fig.~\ref{2D_density_trap}. We observe that a wedding-cake structure appears in the trapped system, and the
coexistence of different phases sensitively depends on the hopping
amplitudes. At lower hopping $t=0.55|U|$, only two phases appear, and the corresponding phase transition is from
a Mott insulator with total filling $n=2$ to a PSF with total filling
$0<n<2$ indicated by the non-vanishing value of $\phi_{bd}$ while $\phi_b = 0$. If we increase the
hopping amplitudes, the first-order tunneling of single atoms will
increase, which induces large density fluctuations in the system, leading to a phase transition from the PSF to the superfluid phase. We clearly observe this point from the middle panel of Fig.~\ref{2D_density_trap} where the superfluid phase starts to
appear right in the middle of the PSF. With further increase of the tunneling
amplitudes the superfluid dominates, as shown in the lower panel of Fig.~\ref{2D_density_trap}, where the PSF completely
disappears at $t=0.7|U|$.

\begin{figure}[h!]
  \vspace{-5pt}
\includegraphics*[width=\linewidth]{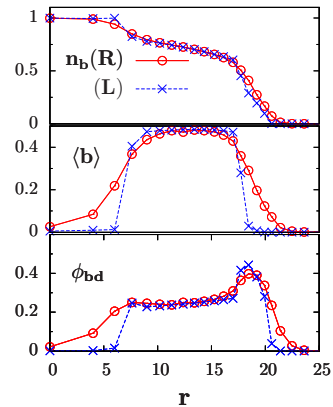}
  \vspace{-30pt}
\caption{Comparison between results from RBDMFT (R) and those from
BDMFT+LDA (L) for a 2D square lattice. Density distributions $n_b$,
order parameters $\phi_b$ and PSF correlator $\phi_{bd}$ vs.
radial distance $r$ at zero temperature in a 2D square lattice. The interactions are $U_b=U_d=2|U|$ and the hopping amplitudes
$t_b=t_d=0.09|U|$ with a harmonic trap $V_0=0.0002|U|$.}\label{2D_density_trap_R_L}
\end{figure}
\begin{figure}[h!]
\vspace{-30pt}
\includegraphics*[width=\linewidth]{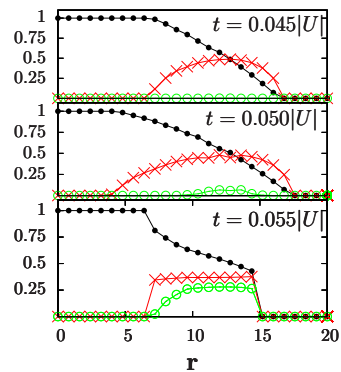}
\vspace{-20pt}
\caption{Density distribution (black line), order parameter
(green line) and PSF correlator $\phi_{bd}$ (red line) vs. radius $r$ for different hopping amplitudes at zero temperature
for a trapped 3D cubic lattice obtained within BDMFT+LDA. The interactions are $U_b=U_d=12|U|$, hopping amplitudes $t=t_b=t_d$ and harmonic trap
$V_0=0.0002|U|$.}\label{density_trap_3D}
\end{figure}
\begin{figure}[h!]
\vspace{-0pt}
\hspace{-10pt}
\includegraphics*[width=3.in]{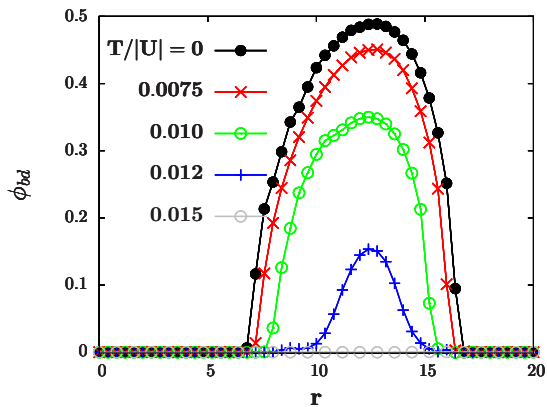}
\vspace{-10pt}
\caption{ Temperature dependence of the PSF correlator
as a function of radius $r$ for a 3D cubic lattice obtained within BDMFT+LDA.
The interactions are $U_b=U_d=12|U|$, hopping amplitudes
$t_b=t_d=0.045|U|$ and harmonic trap
$V_0=0.0002|U|$.}\label{finite_trap_3D}
\end{figure}
Fig.~\ref{2D_density_trap_R_L} shows a comparison between the results of RBDMFT and those of BDMFT+LDA for a 2D square lattice.
We observe good agreement between the two methods except close to the phase transition. In spite of the artificially sharp phase transition feature of LDA, the results of BDMFT+LDA are still reliable with sufficient accuracy in the regime away from the transition. We will therefore apply BDMFT+LDA to tackle the 3D case due to its higher computational efficiency compared to RBDMFT.

Let us now investigate the stability of PSF in the 3D case in the presence of a harmonic trap. Results obtained within LDA are shown in Fig. 8.  Here we choose completely symmetric parameters: $t=t_b=t_d$ and
$U_b=U_d=12|U|$ with balanced filling for the two components.
And only one value for $n_{b,d}$ and $\phi_{b,d}$ is shown
in Fig.~\ref{density_trap_3D}, respectively. Compared to 2D, we observe a similar scenario of phase coexistence
in the trapped 3D cubic lattice: at lower hopping amplitudes the
Mott insulator and the PSF are coexisting, at intermediate hopping amplitudes, the
superfluid phase appears due to increased density fluctuations, and at
even larger hopping amplitudes, the PSF will disappear.

We are also interested in the effects of temperature on the PSF. Fig.
\ref{finite_trap_3D} shows the correlator $\phi_{bd}$ for
different temperatures.  We observe that the PSF is sensitive to
thermal fluctuations. At finite $T$, the PSF is reduced in favor of
developing an unordered phase characterized by $\phi_{bd}=0$. On the other hand, the density distribution is rather insensitive to small finite T.

\subsubsection{Imbalanced mixture}
\begin{figure}[h!]
\vspace{5mm}
\begin{tabular}{ cc }
\hspace{-4mm}
\includegraphics[scale=.52] {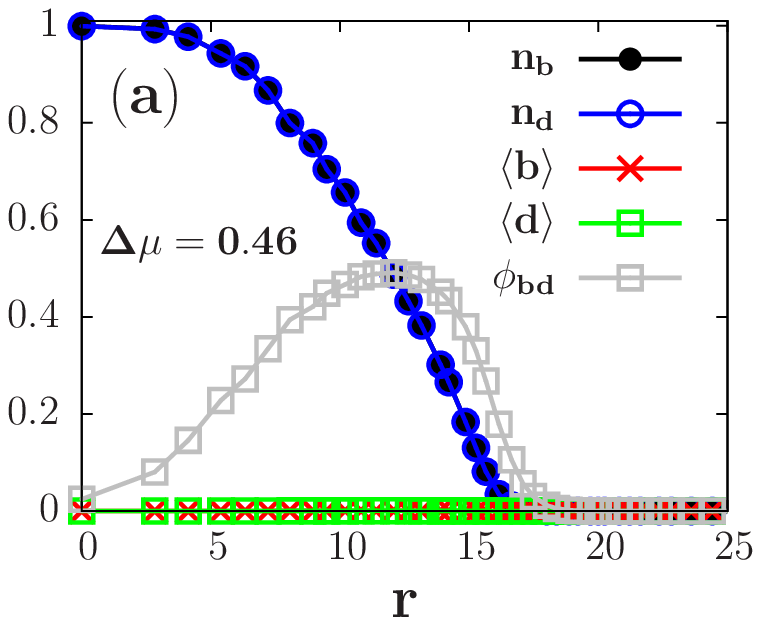}&
\hspace{-8mm}
\includegraphics[scale=.52]{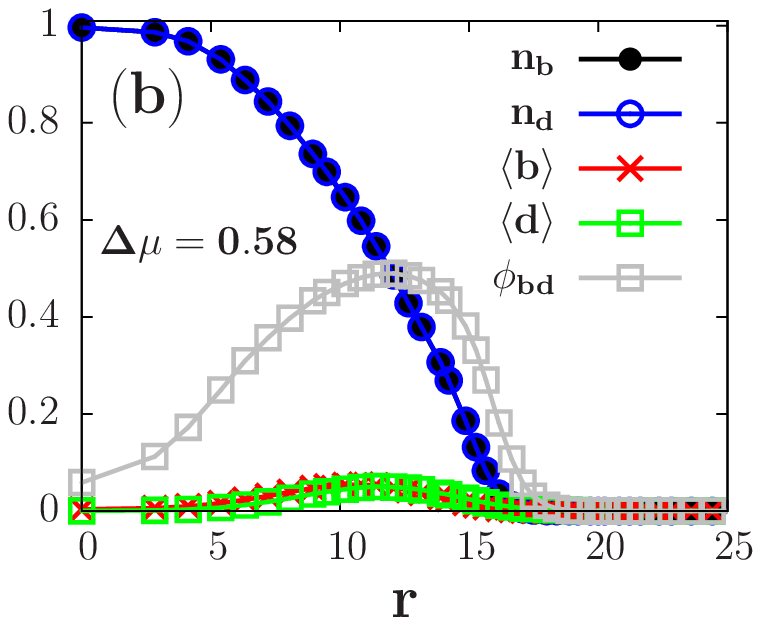}
\\
\hspace{-4mm}
\includegraphics[scale=.52] {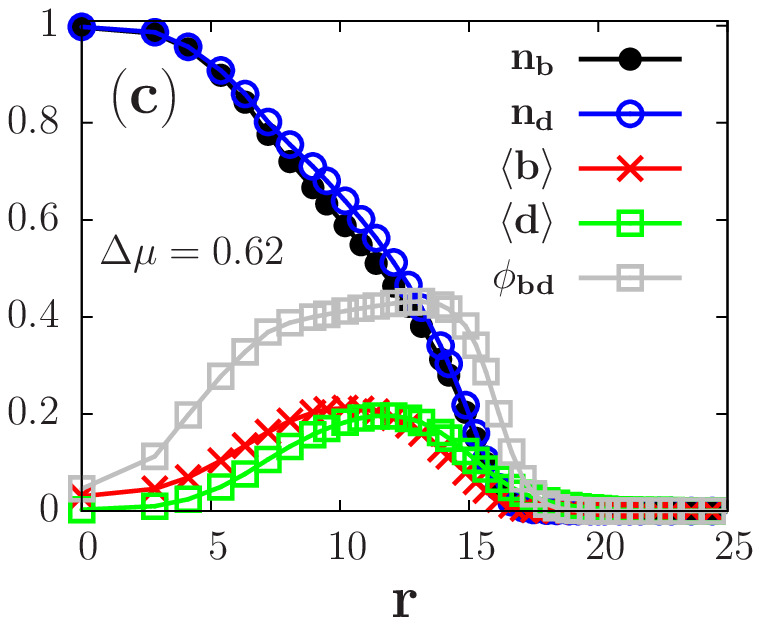}&
\hspace{-8mm}
\includegraphics[scale=.52]{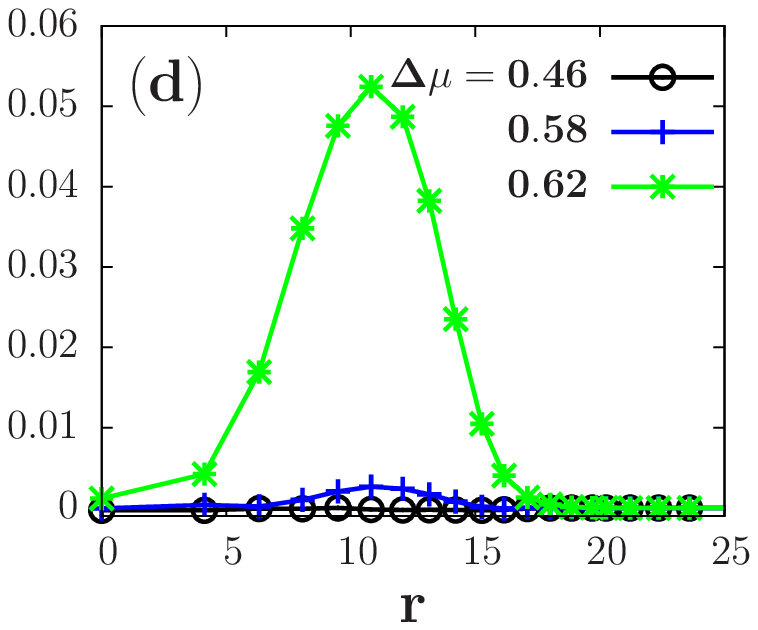}
\end{tabular}
\vspace{-3mm}
\caption{Density distributions $n_{b,d}$, order parameters $\langle
b \rangle$, $\langle d \rangle$ and PSF correlator $\phi_{bd}$ vs.~radial
distance $r$ for different $\Delta \mu$ at zero temperature in a trapped 2D
cubic lattice, obtained within BDMFT+LDA. Panel (d) shows the filling difference
($n_b-n_d$) vs.~radius $r$. The interactions are $U_b=U_d=12|U|$, hopping amplitudes $t_b=t_d=0.05|U|$,
$(\mu_b+\mu_d)/2=0.48$ and harmonic trap $V_0=0.00015|U|$.}\label{Im_density_trap_2D}
\end{figure}
\begin{figure}[h!]
\vspace{5mm}
\begin{tabular}{ cc }
\hspace{-4mm}
\includegraphics[scale=.52] {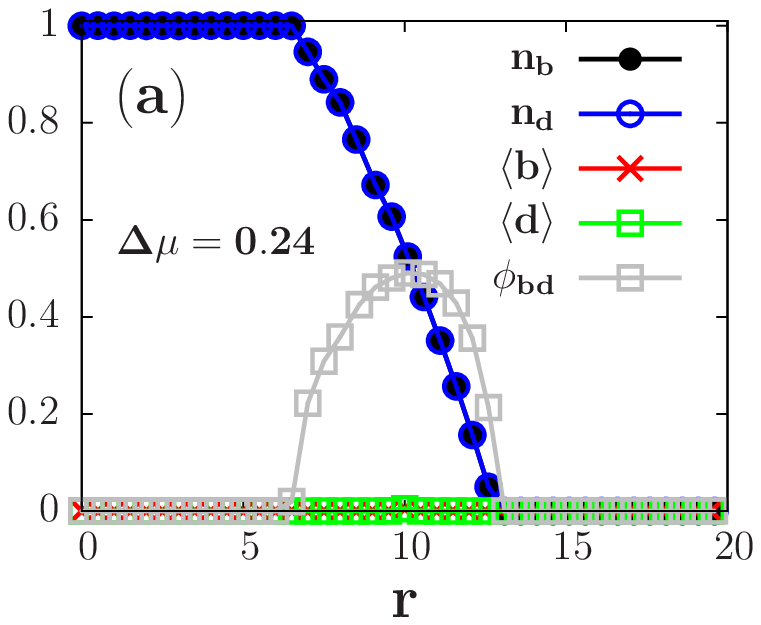}&
\hspace{-8mm}
\includegraphics[scale=.52]{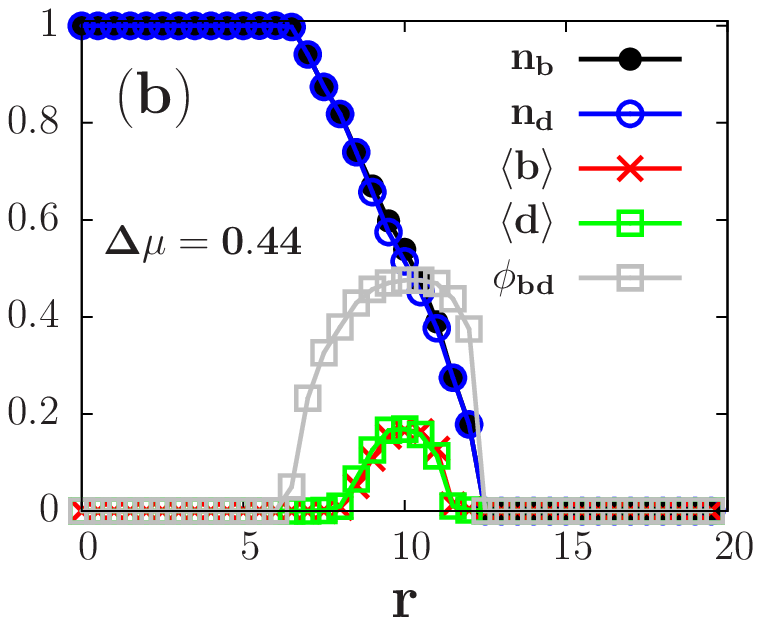}
\\
\hspace{-4mm}
\includegraphics[scale=.52] {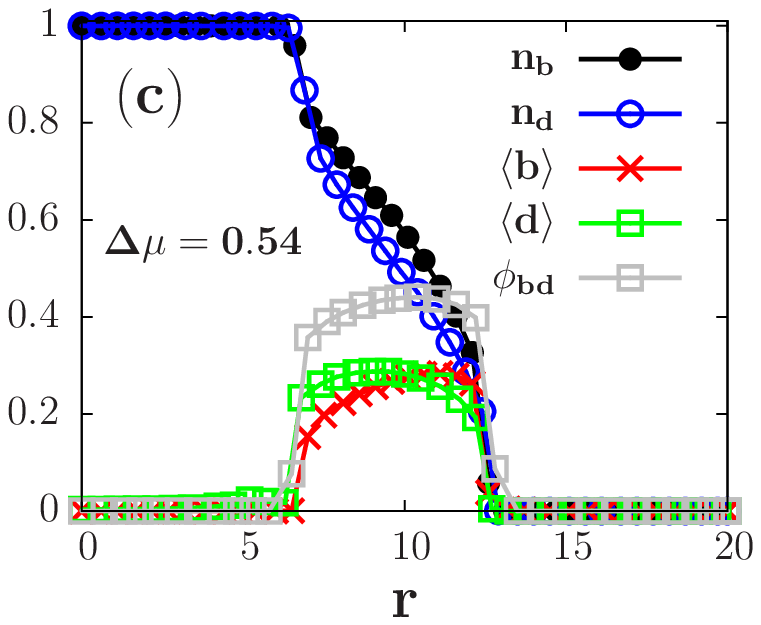}&
\hspace{-8mm}
\includegraphics[scale=.52]{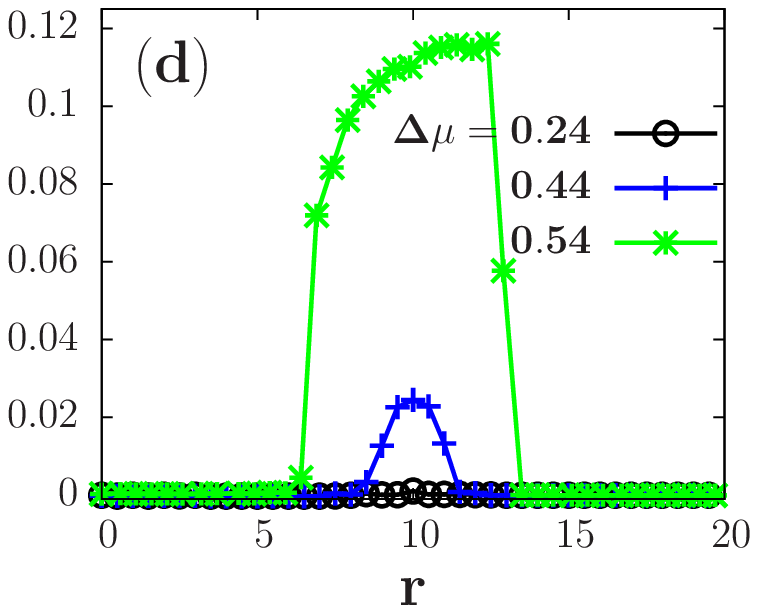}
\end{tabular}
\vspace{-3mm}
\caption{Density distributions $n_{b,d}$, order parameters $\langle b \rangle$, $\langle d \rangle$ and PSF correlator $\phi_{bd}$ vs.~radius $r$ for different $\Delta \mu$ at zero temperature in a trapped 3D cubic lattice, obtained within BDMFT+LDA. Panel (d) shows the filling imbalance ($n_b-n_d$) vs.~radius $r$. The interactions are $U_b=U_d=12|U|$, hopping amplitudes $t_b=t_d=0.04|U|$, $(\mu_b+\mu_d)/2=0.47$ and harmonic trap $V_0=0.0003|U|$.}\label{Im_density_trap_3D}
\end{figure}

As pointed out above, asymmetry in the hopping of the two species does not destroy the PSF.
However, filling imbalance hinders the formation
of the pairs~\cite{C. Menotti_2010, Anzi Hu}. We will now study this effect in more detail. The imbalance, $N_b \neq N_d$, will be controlled
by a nonzero chemical potential difference $\Delta \mu =
\mu_b-\mu_d$ which can be viewed as an effective magnetic field.
Results in 2D are shown in Fig.~\ref{Im_density_trap_2D}. Upon increasing $\Delta \mu$, the PSF will cease to exist and be replaced by a superfluid phase, since the chemical potential difference eventually exceeds the pairing gap for the
PSF, allowing unpaired excess atoms to enter the PSF region.
We find that even small population imbalance can destroy the PSF in a trapped system.
On the other hand, as shown in Fig.~\ref{Im_density_trap_2D}, the density distribution is almost unchanged for similar parameters. When increasing the imbalance parameter $\Delta \mu$ further, the PSF will disappear already for a small population imbalance, since the particles form a conventional superfluid. Here we do not find any phase separation.

Finally we discuss the influence of population imbalance on the PSF in a
trapped 3D cubic lattice using BDMFT+LDA. From our results shown in Fig.~\ref{Im_density_trap_3D} we conclude that here the physics is qualitatively similar to the 2D case.

\section{summary}
We have investigated low-temperature properties of Bose-Bose mixtures with attractive inter-species
interaction both in 2D and 3D optical lattices by means of
BDMFT/RBDMFT. In particular, we found that the pair-superfluid is stable also for asymmetric hopping of the two species. We obtained the critical temperature of the PSF which we found to be of the same order as that of the XY-ferromagnet in the corresponding system with repulsive interactions of equal magnitude. We have confirmed the stability of the PSF in a balanced Bose-Bose-mixture in the presence of the harmonic trap both in 2D and 3D. On the other hand, we found that even a small population imbalance can destroy the PSF. This novel PSF quantum phase can be detected in future experiments via the momentum distribution of pairs, which shows signatures of the pair condensate~\cite{C. Menotti_2010, Anzi Hu}.

\begin{acknowledgments}
We acknowledge useful discussions with  M. Reza Bakhtiari. This work was supported by the China Scholarship Fund (Y.L) and the Deutsche Forschungsgemeinschaft via SFB-TR/49 and the DIP project HO 2407/5-1.
\end{acknowledgments}


\begin{references}
\bibitem{I. Bloch_2008} I. Bloch, J. Dalibard, and W. Zwerger, Rev. Mod. Phys. {\bf 80}, 885 (2008).
\bibitem{I. Bloch_2002} M. Greiner, O. Mandel, T. Esslinger, T. W. H\"ansch and I. Bloch, Nature (London) {\bf 415}, 39 (2002).
\bibitem{J. Catani_2008} J. Catani, L. Desarlo, G. Baronitini, F. Minardi and M. Inguscio, Phys. Rev. A {\bf 77}, 011603(R) (2008).
\bibitem{W. Ketterle_2009} D. M. Weld, P. Medley, H. Miyake, D. Hucul, D. E. Pritchard, and Wolfgang Ketterle, Phys. Rev. Lett. {\bf 103}, 245301 (2009).
\bibitem{W. Ketterle_2010} P. Medley, D. M. Weld, H. Miyake, D. E. Pritchard and W. Ketterle, Phys. Rev. Lett. {\bf 106}, 195301 (2011).
\bibitem{B. Gadway_2010} B. Gadway, D. Pertot, R. Reimann and D. Schneble, Phys Rev. Lett. {\bf 105}, 045303 (2010).
\bibitem{M. Lewenstein_science} M. Lewenstein and A. Sanpera, Science {\bf 319}, 292 (2008).
\bibitem{P. Zoller_2009} A. J. Daley, J. M. Taylor, S. Diehl, M. Baranov and P. Zoller, Phys. Rev. Lett. {\bf 102}, 040402 (2009).
\bibitem{M. Yang_2010} Y. Lee and M. Yang, Phys. Rev. A {\bf 81}, 061604 (2010).
\bibitem{S. Wessel_2011} L. Bonnes and S. Wessel, Phys. Rev. Lett. {\bf 106}, 185302 (2011).
\bibitem{M. Yang_2011} Y. Chen, K. Ng and M. Yang, Phys. Rev. B {\bf 84}, 092503 (2011).
\bibitem{A. Kuklov_1} A. Kuklov, N. Prokof'ev and B. V. Svistunov, Phys. Rev. Lett. {\bf 92}, 030403 (2004); A. Kuklov, N. Prokof'ev and B. Svistunov, Phys. Rev. Lett. {\bf 92}, 050402 (2004).
\bibitem{A. Arguelles_2007} A. Arguelles and L. Santos, Phys. Rev. A {\bf 75}, 053613 (2007).
\bibitem{S. Guertler_2008} S. Guertler, M. Troyer and F. Zhang, Phys. Rev. B {\bf 77}, 184505 (2008).
\bibitem{L. Mathey_2009} L. Mathey, I. Danshita and C. W. Clark, Phys. Rev. A {\bf 79}, 011602 (2009).
\bibitem{A. Hu_2009} A. Hu, L. Mathey, I. Danshita, E. Tiesinga, C. J. Willians and C. W. Clark, Phys. Rev. A {\bf 80}, 023619 (2009).
\bibitem{C. Menotti_2010} C. Menotti and S. Stringari, Phys. Rev. A {\bf 81}, 045604 (2010).
\bibitem{M. Iskin_2010} M. Iskin, Phys. Rev. A {\bf 82}, 033630 (2010); M. Iskin, J. Phys. A {\bf 44}, 275301 (2011).
\bibitem{P. Chen_2010} P. Cheng and M. Yang, Phys. Rev. B {\bf 82}, 180510(R) (2010).
\bibitem{A. Kuklov_2003} A. B. Kuklov and B. V. Svistunov, Phys. Rev. Lett. {\bf 90}, 100401 (2003).
\bibitem{E. Altman_2003} E. Altman, W. Hofstetter, E. Demler and M. D. Lukin, New. J. Phys. {\bf 5}, 113 (2003).
\bibitem{S. Foelling_2007} S. Foelling, S. Trotzky, P. Cheinet, M. Feld, R. Saers, A. Widera, T. Mueller and  I. Bloch, Nature (London) {\bf 448}, 1029 (2007).
\bibitem{S. Trotzky_2008} S. Trotzky, P. Cheinet, S. Folling, M. Field, U. Schnorrberger, A.M. Rey, A. Polovnikov, E.A. Demler, M.D. Lukin and I. Bloch, Science {\bf 319}, 295 (2008).
\bibitem{U. Shrestha_2010} U. Shrestha, Phys. Rev. A {\bf 82}, 041603 (2010).
\bibitem{W. Bakr_2009} W. S. Bakr, J. I. Gillen, A. Peng, S. Foelling and M. Greiner, Nature (London) {\bf 462}, 74 (2009).
\bibitem{J. Sherson_2010} J. F. Sherson, C. Weitenberg, M. Endres, M. Cheneau, I. Bloch and S. Kuhr, Nature (London) {\bf 467}, 68 (2010).
\bibitem{BDMFT} K. Byczuk and D. Vollhardt, Phys. Rev. B {\bf 77}, 235106 (2008); A. Hubener, M. Snoek and W. Hofstetter, Phys. Rev. B {\bf 80}, 245109 (2009); W.-J. Hu and N.-H. Tong, Phys. Rev. B {\bf 80}, 245110 (2009); P. Anders, E. Gull, L. Pollet, M. Troyer and P. Werner, Phys. Rev. Lett. {\bf 105}, 096402 (2010).
\bibitem{RBDMFT} Y.-Q. Li, M. R. Bakhtiari, L. He and W. Hofstetter, Phys. Rev. B {\bf 84}, 144411 (2011); Y.-Q. Li, M. R. Bakhtiari, L. He and W. Hofstetter, Phys. Rev. A {\bf 85}, 023624 (2012).
\bibitem{M. Inguscio_2008} G. Thalhammer, G. Barontini, L. De Sarlo, J. Catani, F. Minardi and M. Inguscio, Phys. Rev. Lett. {\bf 100}, 210402 (2008).
\bibitem{A. Widera_2004} A. Widera, O. Mandel, M. Greiner, S. Kreim, T. W. H\"ansch, and I. Bloch, Phys. Rev. Lett. {\bf 92}, 160406 (2004).
\bibitem{M. Caffarel_1994} M. Caffarel and W. Krauth, Phys. Rev. Lett. {\bf 72}, 1545 (1994).
\bibitem{cooling} D. C. McKay and B. DeMarco, Rep. Prog. Phys. {\bf 74}, 054401 (2011).
\bibitem{RF} C. A. Regal and D. S. Jin, Phys. Rev. Lett. {\bf 90}, 230404 (2003).
\bibitem{Anzi Hu} A. Hu, L. Mathey, C. J. Williams and C. W. Clark, Phys. Rev. A {\bf 81}, 063602 (2010).
\end{references}
\end{document}